\documentclass[11pt]{article}
\usepackage[utf8]{inputenc}
\usepackage{comment}
\usepackage{amsmath,amssymb} 
\usepackage{amsthm}{
 
 \newtheorem{assumption}{Assumption}

}
\usepackage{footnote}
\usepackage{graphicx}
\usepackage{hyperref}

\usepackage{pdflscape}
\usepackage{xcolor}
\usepackage{setspace}

\usepackage{multirow}
\usepackage{array}
\newcolumntype{C}[1]{>{\centering\let\newline\\\arraybackslash\hspace{0pt}}m{#1}}

\title{Comparative Evaluation of Difference in Differences Methods for Staggered Adoption Interventions}

\usepackage{authblk}

\author[1,3]{Ernesto Ulloa-Pérez}
\author[2]{Elizabeth F. Bair}
\author[2,3]{Amol S. Navathe}
\author[1,3]{Kristin A. Linn}

\affil[1]{Department of Biostatistics, Epidemiology and Informatics, University of Pennsylvania}

\affil[2]{Department of Medical Ethics and Health Policy, Perelman School of Medicine, University of Pennsylvania, Philadelphia}

\affil[3]{The Parity Center, Department of Medical Ethics and Health Policy, Perelman School of Medicine, University of Pennsylvania, Philadelphia}

\begin{document}
\maketitle

\begin{abstract}
Staggered adoption is a common approach for implementing healthcare interventions, where different units adopt the program at different times. Difference-in-differences (DiD) methods are frequently used to evaluate the effects of such interventions. Nonetheless, recent research has shown that classical DiD approaches designed for a single treatment start date can produce biased estimates in staggered adoption settings, particularly due to treatment effect heterogeneity across adoption and calendar time. Several alternative methods have been developed to address these limitations. However, these methods have not been fully systematically compared, and their practical utility remains unclear. Motivated by a payment program implemented by a healthcare provider in Hawaii, we provide a comprehensive review of the staggered adoption setting and a selection of DiD methods suitable for this context. We begin with a theoretical overview of these methods, followed by a simulation study designed to resemble the characteristics of our application, where the intervention is implemented at the cluster level. Our results show that the current methods tend to under-perform when the number of clusters is small, but improve as the number of clusters increases. We then apply the methods to evaluate the real-world payment program intervention and offer practical recommendations for researchers implementing DiD methods for staggered adoption settings. Finally, we translate our findings into practical guidance for applied researchers choosing among DiD methods for staggered adoption settings.
\end{abstract}

\section{Introduction}

In many healthcare interventions, a staggered implementation is often necessary due to practical constraints. For example, in 2016, the Hawaii Medical Service Association (HMSA), the Blue Cross Blue Shield of Hawaii, introduced a novel payment system for primary care, which was implemented in a staggered fashion between 2016 and 2019 due to logistical constraints \cite{navathe2019association}. The program's roll-out was not random; physician organizations that adopted the system earlier were selected based on criteria such as geographical representation across the islands and diversity in quality performance.

In order to assess whether the adoption of the HMSA's payment program had an impact on its members, one must consider two key sources of treatment heterogeneity. The first is treatment heterogeneity by time, which occurs when the intervention's effect varies across calendar time. In the HMSA intervention, it may be plausible that the program's effect during the first year after implementation may differ from its effect in the second year. The second is treatment heterogeneity by group, which refers to differences in the effect of the intervention between units that enter the program at different time points. Heterogeneity by group can arise when entry times are associated with the outcome. For instance, a higher percentage of HMSA's health providers that joined in 2016 were urban compared to those who entered the program in 2017 \cite{navathe2019association}. As a result, the effect of the payment program may differ between the groups that adopted in 2016 versus those that adopted it in 2017. 

Effect heterogeneity plays a crucial role in defining the target estimand and selecting an appropriate estimator. Recently, it has been shown that two-way fixed effects regression (TWFE), a widely used method in the difference in differences (DiD) literature, is inadequate for staggered interventions because it does not account for these sources of heterogeneity \cite{goodman-bacon_difference--differences_2021, baker_how_2022, callaway2023difference, de2020two, liu2024practical}. To address this issue, numerous methods have been developed to estimate the effects of staggered interventions, with a focus on carefully defining appropriate estimands and control groups \cite{borusyak24, antonelli2020heterogeneous, callaway2021difference, sun2021estimating, athey2022design, ben-michael_trial_2021, arkhangelsky_synthetic_2021, gardner2022two, callaway2023difference, baker2025difference}. Most of these methods leverage DiD methodology \cite{antonelli2020heterogeneous, callaway2021difference, sun2021estimating, athey2022design}, although other approaches such as target trial emulation \cite{ben-michael_trial_2021}, nested mean models \cite{shahn_structural_2022}, and synthetic controls \cite{arkhangelsky_synthetic_2021} have also been examined. Several studies have been published comparing novel methods that address group-time heterogeneity and we refer to them for in-depth review \cite{goin2023comparing, riddell2023guide, chiu2023and, wang2024advances, roth_whats_2023, Yiqingrev24, wing_designing_2023}. However, few studies have examined methods that account for group-time heterogeneity in a simulation study \cite{wang2024advances, riddell2023guide,ruttenauer2024can}. Moreover, to our knowledge, no studies have examined these methods in a setting with a moderate number of covariates and individuals nested within clusters adopting the intervention, conditions commonly seen in health evaluation contexts, such as our HMSA example.

We carried out a comprehensive literature review and selected four recently-developed estimators for evaluation: DiD for Multiple Time Periods \cite{callaway2021difference}, Interaction-weighted estimator \cite{sun2021estimating}, Two-stage DiD \cite{gardner2022two}, and Two-way Mundlack regression \cite{wooldridge_two-way_2021}. These four methods were selected based on several criteria. Specifically, they account for treatment heterogeneity by time and group in their estimation procedure; are suitable for application to our HMSA data example, which involves limited pre-treatment periods, panel data, and both fixed and time-varying covariates; rely on a DiD framework for estimation, and have publicly available \textsf{R} implementations \cite{rstats}.

Our objective is to provide a comprehensive assessment of the performance of the above DiD methods via a simulation study. Specifically, we aim to address the following questions:

\begin{enumerate}
\item What are the theoretical properties of the estimators? 
\item Using simulated data, how do these methods perform in terms of coverage, mean squared error, and bias?
\item What practical challenges arise when implementing these methods, particularly when treatment is assigned at the cluster level?
\end{enumerate}

Question 1 concerns properties at both the identification and estimation levels. For example: Which types of effects (e.g., group-time specific, event-time, or aggregated) do different methods target? What assumptions are necessary for the estimators to be unbiased? Question 2 has not been thoroughly addressed in the literature. Question 3 arises from practical considerations when data is available at the individual level.

In addition to addressing these questions, we offer practical guidance on selecting estimators based on the results of our simulation study and their theoretical properties. A unique contribution of our simulation study is the evaluation of the impact of cluster-level covariates defined in the treatment assignment, while actual treatment effects are determined by both individual and cluster-level covariates. Finally, we illustrate the performance of the selected estimators using data collected by HMSA to evaluate the effects of its novel payment program on the total number of yearly primary care visits among its member.

This paper is structured as follows. Section \ref{sec:methods} introduces notation and defines the relevant estimands which account for group and time heterogeneity, providing a clear scientific interpretation in the context of staggered adoption interventions. Section \ref{sec:est} offers a theoretical overview of the DiD methods. In Section \ref{sec:sim}, we describe the simulation study as well as the settings used for each method. Sections \ref{sec:res} and \ref{sec:HMSA} present the results of the simulation study and the methods illustration on the HMSA example, respectively. Finally, we conclude with a discussion in Section \ref{sec:disc}, offering practical guidance for applying these methods.

\section{Methods}
\label{sec:methods}

\subsection{Notation and Set-up}

Suppose we observe a sample consisting of $n$ identically distributed observations $O_i$, observed at $\bar{t}$ time points, and grouped within $n_c$ clusters of size $c_j$ where $\sum_{j=1}^{n_c} c_j = n$. Each observation $O_i$ consists of $O_i := \{Y_{i1},\dots,Y_{i\bar{t}}, A_{i1},\dots, A_{i\bar{t}}, X_i, I_{ic}\}$ where $Y_{it} \in \mathbb{R}$ is the observed outcome of individual $i$ at time $t$, $A_{it} \in \{0,1\}$ denotes treatment status at time $t$, $I_{ic}$ is a cluster indicator that is equal to 1 if individual $i$ belongs to cluster $c$ and 0 otherwise, and $X_i \in \mathbb{R}^p$ is p-dimensional a vector of baseline covariates. 

In our setting, $X_i$ may contain information at either the individual-level or cluster-level. For instance, in the HMSA case study, $X_i$ may include both cluster-level characteristics of healthcare providers (e.g., urban/rural classification, type of specialty) or individual-level characteristics of HMSA's members (e.g., age, sex). 


Following the notation used in the methods analyzed, we define the initiation of the treatment intervention at the observation level, rather than at the cluster level. However, in the simulation section, we will be explicit about how treatment is assigned at the cluster level. Let $G_i$ be the first time that observation $i$ receives the intervention: $G_i := \min\{t: A_{it} = 1\}$. If an observation does not receive the intervention we set $G_i=\infty$. Let $G_{ig}$ be the indicator for whether observation $i$ first receives treatment at time $g$ (i.e., $G_{ig} = I(G_i= g$)), and let $\bar{g}$ be the last time at which at least one cluster receives treatment for the first time. Let $\mathcal{G}\subseteq\{2,...,\bar{t}\}\backslash\bar{g}$ denote the support of $G$, which excludes the last to be treated group, or never treated group if available. Additionally, let $C_i = 1(G_i = \infty)$ be the indicator that takes the value 1 if unit $i$ never receives treatment (i.e., those observations whose $G_{ig} = 0$ for all $g \in \mathcal{G}$). The presence of this group in the observed data is not mandatory. 

In order to identify treatment effects, certain staggered adoption methods rely on the probabilities of treatment given covariates, which are a generalization of the propensity score \cite{athey_design-based_2022, callaway2021difference}. We define the unconditional and conditional generalized propensity scores as: $$\bar{\pi}_{g,t}:= P(G_{g} = 1|G_g + (1-A_t)(1-G_g) = 1), \text{ and}$$ $$\pi_{g,t}(x):= P(G_g = 1|X=x, G_g + (1-A_t)(1-G_g) = 1).$$ 
The generalized propensity score represents the conditional probability of initiating the intervention at time $g$ among units that remain untreated at time $g$ or begin the intervention in period $g$. The second version represents the same probability conditional on covariates $X$.

To define the estimands of interest, we use the counterfactual framework. Let $\{Y_{it}(g)\}_{g \in \mathcal{G}}$ denote the set of potential outcomes at time $t$ of observation $i$ had it began treatment at time $g$, and let $Y_{it}(\infty)$ be the counterfactual outcome at time $t$ had it remained untreated.

\subsubsection{Estimands of Interest}

The most granular treatment effect in the staggered adoption setting is: $$\psi_{g,t}:= E[Y_{it}(g) \mid G_g = 1] - E[Y_{it}(\infty) \mid G_g = 1],$$
which represents the contrast in counterfactual means at time $t$ between initializing treatment at time $g$ and never being treated, within the population that began treatment at time $g$. The group-time estimand $\psi_{g,t}$ addresses the question: \emph{within the subpopulation of units that entered treatment in period $g$, what is their average treatment effect at time $t$?} 

Beyond $\psi_{g,t}$, we considered two aggregate estimands that summarize treatment effects over time and groups and which are often of interest to researchers \cite{callaway2021difference}. The first aggregate estimand captures the intervention's effect based on the time elapsed since treatment adoption. Let $\ell$ denote the number of periods since treatment initiation. The effect of treatment $\ell$ periods after adoption is defined as:
\begin{equation}
\label{def:psiell}
\psi_{\ell}:= \sum_{g =1 }^{\bar{g}}\sum_{t = 1}^{\bar{t}} I(t\leq \bar{t})I(t-g = \ell)P(G_g = 1 \mid g+\ell \leq \bar{t})\psi_{g,t}.
\end{equation}
The estimand $\psi_{\ell}$ answers the question: \emph{what is the overall cumulative effect of treatment $\ell$ periods after adoption amongst units who adopted treatment?} 
Note that while the primary interest lies in estimating $\psi_{\ell}$ for $\ell \geq 0$, the estimand is also defined for $\ell < 0$, although it is equal to zero under our assumptions. As discussed in subsequent sections, some methods estimate $\psi_{\ell}$ for negative values of $\ell$, which can be used to assess the validity of the identification assumptions.

The final estimand we considered is a weighted average that combines treatment effects across time and groups to capture the overall intervention effect:
\begin{equation}
\label{def:psiaggr}
\psi_{aggr}:= \sum_{g =1 }^{\bar{g}}\sum_{t = 1}^{\bar{t}} w_{aggr}(g, t) \psi_{g,t}.
\end{equation}
The estimand $\psi_{aggr}$ answers the broader question: \emph{what is the overall effect of the intervention?} The choice of weights $w_{aggr}(g,t)$ depends on the research question. The specific weights used by the methods we evaluated are shown in Table \ref{tab:estimators}. Alternative forms of aggregation of $\psi_{g,t}$ can be considered depending on the scientific objective; see \cite{callaway2021difference} for a comprehensive discussion of these.

\subsection{DiD Methods for Staggered Adoption Interventions}
\label{sec:est} 

We provide a structured overview of the DiD estimators, followed by a summary of each method outlining their key characteristics. For a more detailed overview of these methods, see \cite{callaway2023difference, roth_whats_2023, wang2024advances}. We first introduce four shared assumptions on which these methods are based. 

\subsubsection{Shared Assumptions Across Methods}

\begin{assumption}[Random Sampling] 
The observations $\{Y_{i1},...,Y_{i\bar{t}}, A_{i1},...,A_{i\bar{t}}, X_i\}$ are independent and identically distributed.
\label{assum:1}
\end{assumption}

\begin{assumption}[Irreversibility of Treatment] 
All units are untreated at the first observed time period ($A_1 = 0$). For $t = 2,\dots,\bar{t}$, once a unit receives treatment ($A_{t} = 1$), it remains treated for all subsequent periods ($A_{t+1} = 1$) almost surely.
\label{assum:2}
\end{assumption}
 
\begin{assumption}[Limited Treatment Anticipation] 
There exists a known $\delta \geq 0$ such that
\[
E[Y_{t}(g) \mid X, G_g = 1] = E[Y_{t}(\infty) \mid X, G_g = 1] \quad \text{a.s.}
\]
for all $t \in \{1,\dots,\bar{t}\}$ and $g \in \{1,\dots,\mathcal{G}\}$ such that $t < g - \delta$.
\label{assum:3}
\end{assumption}

\begin{assumption}[Overlap] 
For each $t \in \{1,\dots, \bar{t}\}$ and $g\in \mathcal{G}$, there exists some $\varepsilon >0$ such that
\begin{equation}
P(G_g = 1) > \varepsilon \quad \text{and} \quad \pi_{g,t}(X) < 1 - \varepsilon \quad \text{a.s.}
\end{equation}
\label{assum:4}
\end{assumption} Assumption \ref{assum:1} implies that the dataset consists of repeated measurements over time (i.e., panel data). As noted in Table \ref{tab:methods}, some methods account for additional clustering when estimating the variance, relaxing the independence assumption across units in their uncertainty quantification. Assumption \ref{assum:2} states that all observations are untreated at the initial time period and that once a unit enters treatment, it remains treated for the remainder of the study period. Assumption \ref{assum:3} ensures that, on average, a to-be treated unit's counterfactual outcomes are equivalent $\delta$ periods before treatment initiation. The value of $\delta$ depends on the anticipation assumption required by each method. Specifically, Two Way Mundlack Regression and Two-stage DiD require that $\delta=0$, whereas DiD for Multiple Time Periods and the Interaction Weighted Estimator allow for $\delta>0$. Finally, Assumption \ref{assum:4} guarantees that each unit has a positive probability of starting the intervention at each possible entry time, and that the conditional propensity score remains strictly below one for all covariate values.

\subsubsection{Difference-in-Differences with Multiple Time Periods}

\textbf{Overview:} The method developed by \cite{callaway2021difference} estimates $\psi_{g,t}$ using weighting, regression, or a combination of both. Intuitively, this approach fits conditional DiD regressions within each group and time, using carefully selected untreated units as controls. For each $(g,t)$, the method estimates a parametric model for the nuisance functions defined below based on the regression method and control group selection. The estimated nuisance functions are then plugged into the sample analogue of $\psi_{g,t}$. Finally, the method aggregates the estimated $\hat{\psi}_{g,t}$ to obtain estimates of $\psi_{\ell}$ and $\psi_{aggr}$, using the weights described in Table \ref{tab:methods}. \textbf{Control groups:} This method allows the user to select either the never-treated units or the not-yet-treated group (which include the never-treated units) as controls. \textbf{Additional identification assumptions:} If using the not-yet-treated group as controls, the estimator requires the following parallel trends assumption:

\begin{assumption}[Parallel Trends in Difference-in-Differences with Multiple Time Periods] 
For each $g \in \mathcal{G}$ and each $(s,t) \in \{2,....\bar{t}\} \times\{2,...,\bar{t}\}$ such that $t\geq g-\delta$ and $t+\delta \leq s$, where $\delta$ is the treatment anticipation time from Assumption \ref{assum:3}, it holds that
\begin{equation}
E[Y_t(\infty) - Y_{t-1}(\infty) \mid X, G_g = 1] = E[Y_{t}(\infty) - Y_{t-1}(\infty) \mid X, A_s = 0, G_g = 0].
\label{eq:assumdidCS}
\end{equation}
\end{assumption}
The above assumption, often referred to as \emph{conditional parallel trends}, generalizes the standard parallel trends assumption from the two-period difference-in-differences setting. If the never-treated group is used as controls, a slightly modified version of Equation \eqref{eq:assumdidCS} is required, where the conditioning population on the right-hand side consists of the never-treated units. Additionally, if treatment anticipation is expected, $\delta$ can be set to a value greater than zero.

To estimate treatment effects, this method uses parametric models for the difference in conditional means of $Y$ across time given $X=x$: 
\[
m_{g,t}(x) := E[Y_t - Y_{g-\delta -1} \mid X =x, A_{t+\delta} = 0, G_g = 0],
\]
or the generalized propensity score given $X=x$, $\pi_{g,t}(x)$, which was defined in the previous section.

A key strength of this method is its \emph{double robustness} property: it provides consistent estimates of $\psi_{g,t}$ if either the parametric model for $m_{g,t}$ or $\pi_{g,t}$ is correctly specified. \textbf{Limiting distribution:} Under additional parametric and smoothness assumptions pertaining to the nuisance functions, the asymptotic normality of this estimator is derived in \cite{callaway2021difference} (Theorem 2). Variance estimation is performed using the bootstrap, which can account for clustering when resampling is performed a the cluster level (Remark 10).

\subsubsection{Interaction-weighted Estimator}

\textbf{Overview:} The estimator by \cite{sun2021estimating} first targets a conditional estimand $\psi_{g,g+\ell}$ and then marginalizes these to obtain weighted versions of $\psi_{\ell}$ and $\psi_{aggr}$. Broadly, this method estimates the outcome as a function of event-time indicators (and other relevant adjustment covariates) to then find marginal estimates of event-time and aggregated effects. The interaction-weighted estimator follows three main steps:
\begin{enumerate}
\item Estimate the treatment effects of duration $\ell$ within each group $g$, denoted as $\alpha_{g,\ell}$, via the regression model:
 \begin{equation}
 Y_{it} = \eta_i + \mu_t + \sum_{g \notin \mathcal{C}}\sum_{\ell \neq -1} \alpha_{g, \ell} (I\{G_{ig} = 1\} D_{it}^\ell) + \epsilon_{it}.
 \label{eq:iwSA}
 \end{equation}
Here, $D_{it}^{\ell} = I(t-g_i = \ell)$ is an indicator for relative period $\ell$, $\eta_i$ and $\mu_t$ are individual and time fixed effects, respectively, and $\epsilon_{it}$ is an error term that allows for correlation across time. In this model, $\alpha_{g, \ell}$ is a conditional estimate of the intervention effect for group $g$ after $\ell$ exposure periods. The sum of each exposure indicators in Equation \ref{eq:iwSA} is over all groups and lengths of exposure times, except for group $\mathcal{C}$ and length equal to -1, which automatically sets the control group to be observations in $\mathcal{C}$. 

 \item Estimate the weights $\mathrm{P}(G_g = 1 \mid g \in [-\ell, \bar{t}-\ell])$, which represent the sample share of each cohort that experiences at least $\ell$ relevant periods relative to treatment initiation.
 \item Combine the estimated weights and treatment effects to obtain an aggregate effect across specified treatment durations $\mathcal{L}$:
 \begin{equation}
 \frac{1}{|\mathcal{L}|} \sum_{\ell \in \mathcal{L}} \sum_{g} \hat{\alpha}_{g,\ell} \hat{\mathrm{Pr}}(G_g = 1 \mid g\in [-\ell, \bar{t}-\ell]).
 \label{eq3:iwSA}
 \end{equation} If $\mathcal{L}$ is a singleton $\ell$, the method estimates $\psi_{\ell}$:
 \begin{equation}
 \hat{\psi_{\ell}} = \sum_{g = 1}^{\bar{g}} \hat{\alpha}_{g,\ell} \hat{\mathrm{Pr}}(G_g = 1 \mid g\in [-\ell, \bar{t}-\ell]).
 \label{eq4:iwSA}
 \end{equation} If $\mathcal{L}$ includes all non-negative treatment durations (denoted by $\mathcal{L}^+$), the estimator targets $\psi_{aggr}$.
\end{enumerate}
\noindent \textbf{Control groups:} The reference group usually consists of the last-treated cohort (or the never-treated cohort, if available) and contrasts are computed relative to the period prior to treatment ($\ell = -1$).
\noindent \textbf{Additional identification assumptions:} This method requires the following parallel trends assumption:
\begin{assumption}[Parallel Trends for Interaction-weighted Estimator] 
For all $s \neq t$,
\[
E[Y_{t}(\infty) - Y_{s}(\infty) \mid G_g =1]
\]
is the same for all $g \in \mathcal{G}$, where $\mathcal{G}$ is the support of $G$.
\label{asum:sunandab}
\end{assumption}
If this assumption holds and there is no treatment anticipation, $\hat{\alpha}_{g, \ell}$ is a consistent estimator of $\psi_{g, g+\ell}$. The method can also relax the parallel trends assumption to its conditional version on covariates \cite{abadie_semiparametric_2005, sun2021estimating}. \textbf{Limiting distribution:} The estimator is asymptotically normal under additional assumptions regarding the parametric model in Equation \ref{eq:iwSA} (Assumption 4 in \cite{sun2021estimating}), such as finite fourth moments of the error terms. A closed-form of the variance is provided \cite{sun2021estimating} which can account for one level of clustering.  

\subsubsection{Two-stage Difference-in-Differences}

\textbf{Overview:} The method proposed by \cite{gardner2022two} estimates $\psi_{\ell}$ and $\psi_{aggr}$ using a two-step regression approach. Concisely, this estimator imputes untreated counterfactuals for treated units using a linear model that adjusts for time trends and covariates, then estimates treatment effects by comparing the observed and imputed outcomes. The first stage removes time and group trends using untreated observations, while the second stage estimates the intervention's effect. Mathematically, recall that $A_{it}$ is the  indicator of treatment at time $t$, and let $\tilde{Y}_{it}$ denote the transformed outcome where the individual fixed effects are removed: $\tilde{Y}_{it}:= Y_{it} - T_i^{0}\sum_{t=1}^T Y_{it}$ where $T_i^{0}$ is the last time before an individual gets treated $T_i^0:= \sum_{t=1}^T(1-D_{it})$. Define $\tilde{X}_i$ in a similar manner.

In the first stage, among the untreated observations ($A_{it} = 0$), the method regresses the transformed outcomes on group, time, and transformed covariates:
\[
\tilde{Y}_{it} = \mu_g + \mu_t + \beta'\tilde{X}_i + \epsilon_{igt}.
\]
The estimated time and group effects are then removed to obtain residualized outcomes:
\[
\tilde{Z}_{it} := \tilde{Y}_{it} - \hat{\mu}_g -\hat{\mu}_t - \hat{\beta}'\tilde{X}_i.
\]
For the second stage, the residualized outcome $\tilde{Z}_{it}$ is regressed on treatment status $A_{it}$:
\[
\tilde{Z}_{it} = \gamma A_{it} + \eta_{it}.
\]
The coefficient $\gamma$ provides an estimate of $\psi_{aggr}$, and the choice of observations in the second-stage regression determines its weights. Using all available treated observations, the estimator targets:
\[
\sum_{g = 1}^{\bar{g}}\sum_{t = g}^{\bar{t}} P(G = g, T = t \mid D_{gt} = 1) \psi_{g,t},
\]
whereas excluding observations with treatment durations longer than $\bar{\ell}$ targets the $\bar{\ell}$-period average:
\[
\sum_{g = 1}^{\bar{g}}\sum_{t = g}^{g+\bar{\ell}-1} \left[ P(G = g \mid D_{gt} = 1)/\bar{\ell} \right] \psi_{g,t}.
\]
Event-time effects $\psi_{\ell}$ can be estimated by regressing $\tilde{Z}_{it}$ on $D_{it}^{\ell}$. 
\textbf{Control Groups:} The control units in the first stage consist of all observed untreated units. \textbf{Additional identification assumptions:} This method assumes the following parallel trends assumption.
\begin{assumption}[Parallel Trends for Two-stage DiD] For all $g$ and $it$, there exists non-stochastic $\gamma$, $\lambda_i$, $\alpha_t$ such that 
\label{asum:2sdid}
\begin{equation*}
E[Y_{it}(\infty)|X_{i}, G_i=g] = \lambda_i+ \alpha_t + \gamma^\top X_{i}.
\end{equation*}
\end{assumption}
Similar to Assumption \ref{asum:sunandab}, the above assumption assumption imposes a linear structure on the never-treated counterfactual outcomes for each group and time periods. This assumption includes parallel trends during the pre and post periods, whereas Difference-in-Differences with Multiple Time Periods requires parallel trends after the initiation treatment times. Additional regularity assumptions are necessary to derive the asymptotic distribution of the estimator. 
\textbf{Limiting Distribution:} The asymptotic distribution of the second-stage estimates follows from modeling the two-stage procedure as a joint generalized method of moments estimator \cite{hansen1982large}. Current software conducts variance estimation using the  bootstrap \cite{did2s}.

\subsubsection{Two-way Mundlack Regression}

\textbf{Overview:} The Two-way Mundlack regression method \cite{wooldridge_two-way_2021} primarily targets $\psi_{g,t}$; however, software implementations have been developed to estimate versions of $\psi_{\ell}$ and $\psi_{aggr}$ \cite{McDermott}. Concisely, this approach fits a single regression model (with the aim of re-adjusting TWFE) with carefully selected covariates to account for group and time heterogeneity, then marginalizes coefficients to obtain unconditional treatment effects.

The method first estimates the conditional treatment effect given covariates:
$\psi_{g,t}(x):= E[Y_t(g) - Y_t(\infty) \mid X = x, G_g = 1].$
where $\psi_{g,t}(x)$ is a linear function of the outcomes on covariates, group and time indicators, and carefully defined interaction terms to account for treatment heterogeneity across time and groups. After estimating the conditional treatment effects, marginal estimates $\psi_{g,t}$ can be obtained by integrating $\psi_{g,t}(X)$ over the distribution of covariates $X$. Similarly, estimates of $\psi_{aggr}$ and $\psi_{\ell}$ can be obtained by marginalizing over $X$ and $T$ (for $\psi_{aggr}$) or over different treatment durations $\ell$ (for $\psi_{\ell}$). 

The weights $w_{\ell}(g,t)$ and $w_{aggr}(g,t)$ (as defined in Equations \ref{def:psiell} and \ref{def:psiaggr}) are not explicitly specified in this method but instead depend on the empirical distributions of $G_g$, $T$, and $X$. The linear model used to estimate $\psi_{g,t}(x)$ does not include parameters for past treatment effects, making it infeasible to estimate $\psi_{\ell}$ for $\ell < 0$. \textbf{Control Groups:} All untreated observations are used as controls in the regression model. \textbf{Additional Identification Assumptions:} In addition to Assumptions \ref{assum:1}-\ref{assum:4}, this method requires the following \emph{conditional common trends} assumption:
\begin{equation}
\label{asum1:CS}
E[Y_t(\infty) - Y_1(\infty) \mid G_1, \dots, G_{\bar{t}}, X] = E[Y_t(\infty) - Y_1(\infty) \mid X].
\end{equation}
which states that, conditional on covariates $X$, the expectation of the difference in untreated counterfactuals between times $t$ and $1$ does not depend on group indicators $G_1,...,G_{\bar{t}}$. To estimate the conditional treatment effect $\psi_{g,t}(x)$, the conditional mean $E[Y_t(\infty) \mid G_1, \dots, G_{\bar{t}}, X]$ must be correctly specified (see Equations 6.27–6.29 in \cite{wooldridge_two-way_2021} for details). Specifically, this mean function must be linear in covariates, time indicators, treatment indicators, and interactions between covariates with time and group indicators. Additionally, this method requires that the limited treatment anticipation assumption (Assumption \ref{assum:3}) holds with $\delta = 0$, meaning that units do not adjust their outcomes in anticipation of treatment. \textbf{Limiting Distribution:} The estimator $\hat{\psi}_{g,t}(x)$ is unbiased and consistent as long as there are no collinear covariates and each exposure group has a sufficient sample size as $n \to \infty$. Fixing the number of time periods, $\hat{\psi}_{g,t}(x)$ is asymptotically normal, and by the delta method, so are the estimates of $\psi_{g,t}$, $\psi_{\ell}$, and $\psi_{aggr}$ \cite{marginaleffR}.

\begin{table}
\begin{center}
\begin{tabular}{|C{6em}|C{18em}|C{18em}|} \hline 
\textbf{Method} & $w_{\ell}(g,t)$& $w_{aggr}(g,t)$\\ \hline 
Diff-in-Diff for multiple time periods & $I(t\leq \bar{t})I(t-g = \ell)P(G_g = 1 |g+\ell \leq \bar{t})$& $I(t\geq g)P(G=g | G\leq \bar{t})/(\bar{t}-g+1)$\\ \hline
Interaction Weighted Estimator & $I(t-g = \ell) \mathrm{P}(G_g = 1| 0 \leq g + \ell \leq \bar{t})$ & $\frac{1}{|\mathcal{L}^+|}I(t-g = \ell) \mathrm{P}(G_g = 1|0 \leq g + \ell \leq \bar{t})$ \\ \hline
Two-stage Diff-in-Diff & Unavailable & $P(G = g, T = t| D_{gt} = 1)I(t \leq \bar{t})$ \\ \hline
Two-way Mundlack & Depend on distribution of $G_g$ and $X$ &Depend on distribution of $G_g$ and $X$ \\ \hline
\end{tabular}
\end{center}
\caption{Summary of target estimands and corresponding weights for each method considered. Event-time weights for Two-stage DiD are not provided in \cite{gardner2022two}. Weights for Two-way Mundlack Regression depend on the distribution of $G_g$ and $X$ due to the marginalization step and therefore have no closed-form solution. Additional options for different weights exist for Two-stage Difference-in-Differences (DiD) and DiD with multiple time periods.}
\label{tab:estimators}
\end{table}

\subsection{Simulation Study}
\label{sec:sim}

\subsubsection{Data generating mechanism}

We evaluated the methods on various simulation scenarios in which we varied the total number of clusters, sample size, and the effect of exposure on the outcome. The total sample sizes in the scenarios were 500, 1000, 2000, and 5000. The number of clusters also changed within each setting and was equal to 30, 50, and 100. The number of individuals per cluster was equally distributed. The effect of exposure on the outcome was modeled under three scenarios: constant (no heterogeneity), lagged (temporal heterogeneity), and heterogeneity by both group and time. The total number of time points was set to $\bar{t}=5$, and we enforced that each cluster start the intervention between periods 3 and 5, i.e. $G_g =0$ for $g = 1,2$. Hence, there were no never-treated units and two observed periods with no intervention effect. 

We generated five covariates in total, three at the individual level and two at the cluster level. Among the three individual covariates, two were continuous and one binary. These were meant to resemble our motivating example, in which the individual-level covariates are age, a frailty score, and sex. The cluster-level covariates consisted of one categorical covariate with 3 levels and one binary covariate. Similarly, these covariates were meant to mimic clinic characteristics such as type of practice, and whether the clinic is located in an urban or rural area. 

To create an association between individual-level covariates and clusters, we first generated cluster-level means, which were then used as parameters for generating the individual-level covariates. The individual-level covariates were drawn from the following distributions:
$$X_1|\mu_c \sim N(\mu_{c},1),$$
$$X_2|\nu_c \sim N(\nu_{c},1),$$
$$X_3|p_c \sim \text{binary}(p_c),$$
where $\mu_{c} \sim N(5,4), \nu_{c} \sim N(5,1)$ and $p_{c} \sim \text{Uniform}(0.5,0.6)$ were generated for each cluster $c$. The cluster-level covariates had the following distribution
$$X_{4} \sim \text{multinomial}(0.4, 0.3, 0.3), \text{ and} $$
$$X_{5} \sim \text{binary}(0.7).$$
The generalized propensity score of beginning treatment at a given time point depended on cluster-level covariates exclusively. This dependence was meant to resemble the HMSA example, where the probabilities of entering the payment program depend exclusively on the provider. We generated the probabilities of each of the entry times $3,4$ and 5 with a multinomial distribution, with probability parameters for each time given by
$$C_k:= \frac{\text{exp}\left(\theta_k^{\top}X\right)}{\sum_{t=1}^T\text{exp}\left(\theta_t^{\top}X\right)}$$
where $k \in \{3,4,5\}$. Because $\log(C_k/C_j) = \theta_k^{\top}X -\theta_j^{\top}X = (\theta_k-\theta_j)^{\top}X$, a multinomial model to fit the estimated probabilities of entry would be correctly specified. The parameters $\theta_3,\theta_4,\theta_{5}$ were defined such that the probabilities of beginning the intervention at the different time points were larger throughout time. That is, $C_3<C_4<C_5$. Information on the coefficients and the marginal probabilities of entry can be found in the Appendix (Tables \ref{tab:margprobs} and \ref{tab:coefssim}). 

\begin{figure}[h!]
\centering
\includegraphics[scale=0.23]{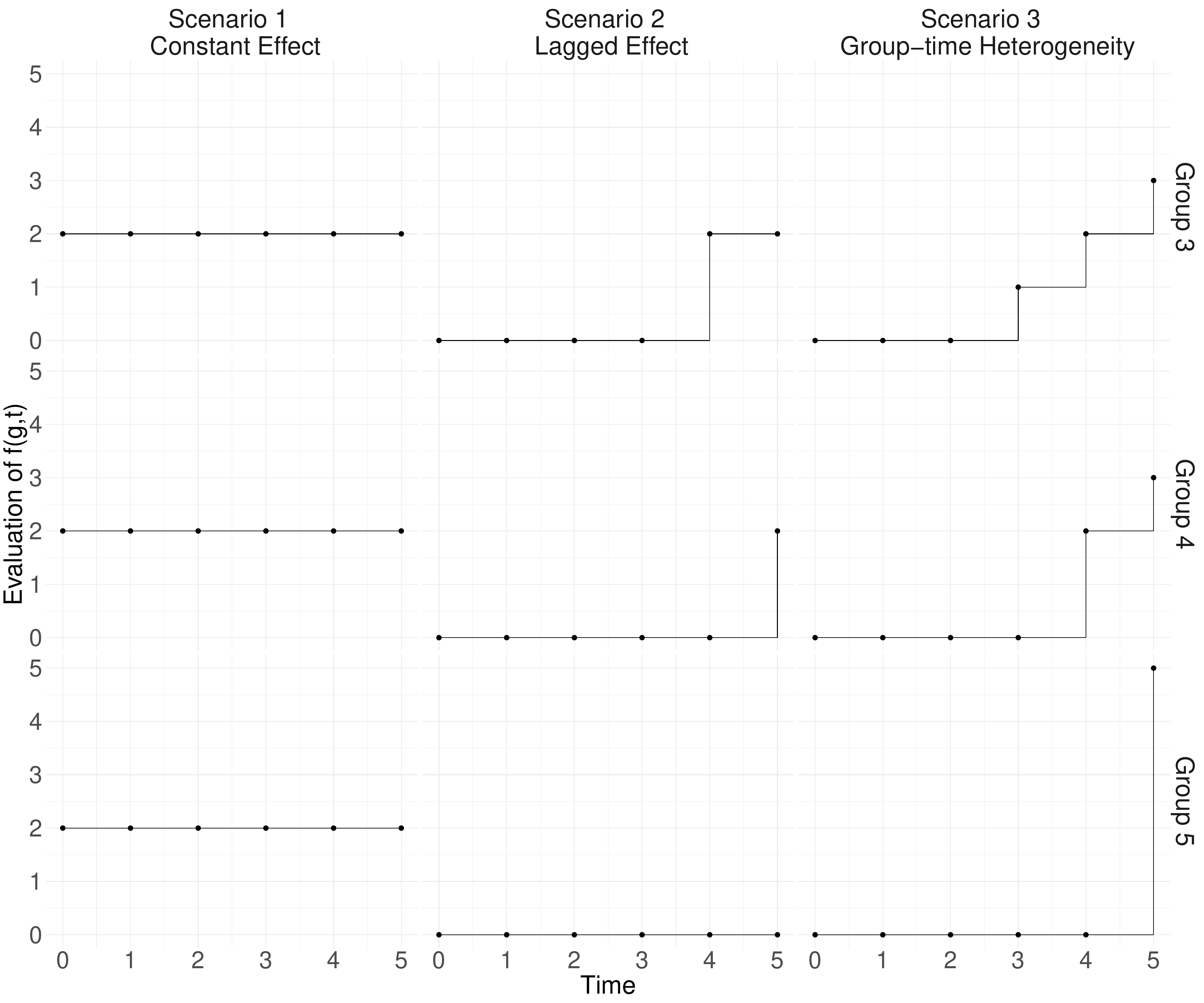}
\caption{Evaluation of $f(g,t)$ across all combinations of $g$ and $t$. Each column corresponds to a simulation scenario, and each row represents a treatment group $G$.}
\label{fig1:grptime}
\end{figure}

The outcome was modeled using individual and cluster-level covariates, time trends, individual-level effects, and random effects at the cluster-level. To introduce confounding, the coefficients of the individual and cluster-level covariates varied over time \cite{zeldow2021confounding}. The untreated counterfactual outcome in cluster $c$ was generated as: 
\begin{equation}
Y_{itc}(\infty) = \mu_t + X_i^\top \beta_t + \eta_i + \xi_{c} + \epsilon_{it},
\end{equation}
where $\mu_t = t$, $\beta_t$ was set to $1 + (t-1)/5$ at time $t$ for each variable in $X$, $\eta_i \sim N(0, 0.5)$ and $\xi_{c} \sim N(0, 0.5)$ are the individual and cluster-level random effects, respectively, and $\epsilon_{it} \sim N(0, 0.5)$. 
The treated counterfactual had a similar form, except it also included a function that depended on group and time:
\begin{equation}
Y_{itc}(g) = I(g \leq t ) f(g,t) + \mu_t + X_i^\top \beta_t + \eta_i + \xi_{c} + \epsilon_{it}
\end{equation}
Different forms of $f(g,t)$ were specified for each scenario. In Scenario 1, the treatment effect was constant over time. In Scenario 2, the effect was lagged, initiating after the second exposure period. In Scenario 3, the treatment effect was heterogeneous across both time and group. Figure \ref{fig1:grptime} illustrates the values of $f(g,t)$ for all possible $(g,t)$ combinations across each scenario.

\subsubsection{Performance metrics and method specifications}

All methods were evaluated with respect to their empirical mean squared error (MSE), absolute bias, and empirical 95\% coverage. We carried out 1,000 simulations for each setting. 

We fit the staggered adoption methods on the simulated datasets using \textsf{R} statistical software, version 4.2.1. We implemented both the weighted and doubly robust approaches for the DiD estimator for multiple time periods using the \textsf{R} package \textsf{did}. Both approaches utilize observations that were not yet treated as controls and included all available covariates. Although the generalized propensity score approach does not include all covariates, we included all of them in the weighted approach to mimic what would be done in practice. For the interaction-weighted estimator, we used the \textsf{sunab} function from the \textsf{fixest} package, which adjusts for covariates by demeaning them from the dataset. To fit the two-stage difference-in-differences method, we used the \textsf{did2s} package, specifying all covariates in the first stage of the regression model. Finally, we used the \textsf{etwfe} package to fit the Two-way Mundlack regression and adjusted for all covariates. The code for fitting the methods is available on Github \cite{ulloa2025staggeredcode}. All methods accounted for correlation at the cluster level when computing their standard errors. Table \ref{tab:methods} provides an overview of the information used to fit the methods.

\begin{table}
\begin{center}
\begin{tabular}{|C{6em}|C{10em}|C{6em}|C{5em}|C{8em}|C{6em}|} \hline 
\textbf{Method}& \textbf{Nuisance functions} & \textbf{Target Estimands} & \textbf{Controls} & \textbf{Clustering} & \textbf{\textsf{R} package, function} \\\hline 
Diff-in-Diff multiple time periods, doubly robust & $E[Y_t - Y_{g -1}|X, A_{t} = 0, G_g = 0]$ and 
$\pi_{g,t}(x)$& $\psi_{g,t}, \psi_{\ell}, \psi_{aggr}$& Not yet treated & Yes, cluster size needs to be large for bootstrap to work & did, att\_gt \\ \hline
Diff-in-Diff multiple time periods inverse probability weights & 
$\pi_{g,t}(x)$ & $\psi_{g,t}, \psi_{\ell}, \psi_{aggr}$ & Not yet treated & Yes, cluster size needs to be large for bootstrap to work & did, att\_gt \\ \hline
Interaction Weighted Estimator & $E[Y_{it}| \alpha_i, \mu_t, G_g D_{it}^\ell, X]$ & $ \psi_{\ell}, \psi_{aggr}$& Last to be treated group & Covariance matrix can account for clustering or individual-level correlation, but not both$^1$. & fixest, feols\\ \hline
Two-stage Diff-in-Diff & $E[Y_{igt}|\mu_g,\mu_t, X]$& $\psi_{\ell}$, $\psi_{aggr}$& All untreated units& Yes, software carries bootstrap at the cluster level &did2s, did2s \\ \hline
Two-way Mundlack Regression & $ E[Y_{it}|X, B_{gt}, G_g ]$ &$\psi_{g,t}, \psi_{\ell}, \psi_{aggr}$ & All untreated units &Linear model to estimate $\psi_{g,t}(x)$ can account for clustering, but marginalization software currently does not & etwfe, etwfe \\ \hline
\end{tabular}
\end{center}
\caption{Summary of methods characteristics and their settings utilized in the simulation study. $^1$Correlation was chosen at the cluster level in this simulation study.}
\label{tab:methods}
\end{table}

\section{Results}
\label{sec:res}

\subsection{Simulation results}

Because both event-time and marginal estimands depend on weights determined by the data-generating mechanism, the methods evaluated may target different estimands. This distinction is particularly important in Scenario 3, where the treatment effect varies across both time and group; in this setting, the marginal estimands depend on the weighting scheme specific to each method. Table \ref{tab:avgestimates} in the Appendix illustrates the observed differences of event time and marginal estimated values across scenarios and methods.

To ensure a fair comparison, we focused on settings where the estimand is invariant to the weighting scheme. Specifically, we present results for estimation of: group-time effects in Scenario 3, when $f(g,t)$ varies across both group and time; event-time effects in Scenario 2, where the effect is lagged by one time period; and marginal effects in Scenario 1 where the effect is constant across time and groups. 

\subsection{Group-time effect results}

\begin{figure}[h!]
\centering
\includegraphics[scale=0.28]{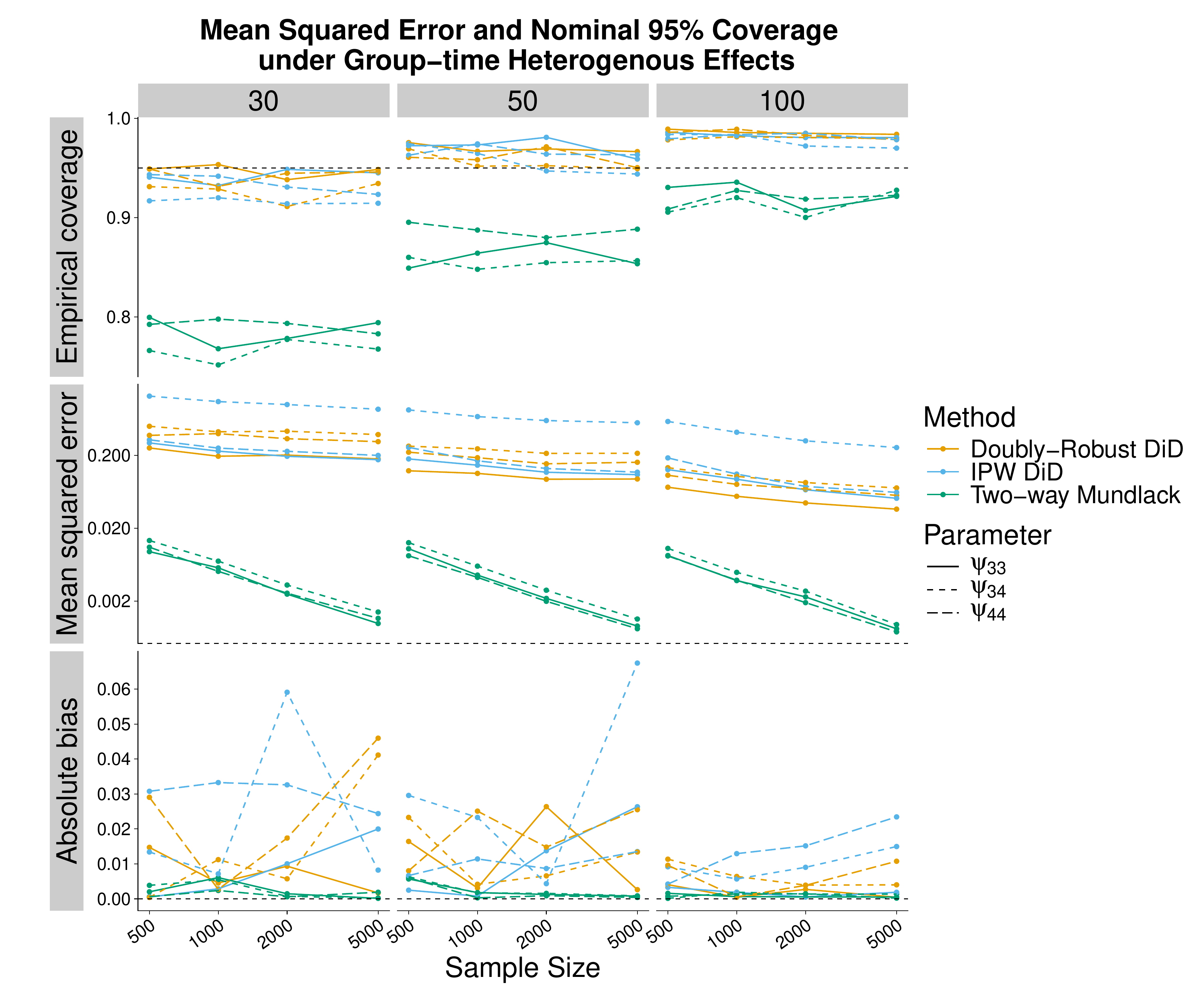}
\caption{Performance results by sample size and number of clusters for Scenario 3 when estimating $\psi_{3,3}$, $\psi_{3,4}$, and $\psi_{4,4}$. Top panel: empirical 95\% coverage. Middle panel: mean squared error (MSE). Bottom panel: absolute bias. The weighted DiD for multiple time periods had the highest MSE and absolute bias in most cases, while the Two-way Mundlack regression had the lowest across all settings. The doubly robust DiD estimator showed decreasing MSE and decreasing bias with increasing number of clusters, and coverage close or above nominal across all settings. All approaches had lower than nominal coverage with few clusters, but coverage increased as the number of clusters increased.}
\label{fig2:grptime}
\end{figure}

In Scenario 3 with group and time heterogeneity, coverage, mean MSE, and bias varied by method. Overall, all metrics improved as the number of clusters increased (Figure \ref{fig2:grptime}). Specifically, Two-way Mundlack regression had coverage below the nominal 95\% across all sample sizes and number of clusters. When the number of clusters was 30, its coverage ranged from 60\% to 85\%, and improved with increasing sample and cluster size. However, even in the largest sample settings, it did not reach 95\% nominal coverage. The doubly robust DiD and IPW estimators showed under-coverage in settings with 30 clusters, and presented over-coverage as the number of clusters increased. The over-coverage is consistent with the method’s approach to estimating multiple parameters, where simultaneous confidence intervals are designed to achieve at least 95\% coverage. MSE and bias also improved as number of clusters increased, however, bias did not necessarily improve as sample size increased.

Across all estimands $\psi_{g,t}$, Two-way Mundlack regression had lower MSE than both the doubly robust and IPW DiD for multiple time periods approaches. For all methods, coverage and precision were better for $\psi_{3,3}$ than for $\psi_{3,4}$, due to the greater availability of control observations at $t = 3$. Precision for $\psi_{3,4}$ was also higher than for $\psi_{4,4}$ for the same reason. Results were similar in scenarios where the effect was lagged and constant (Appendix Figures \ref{fig1app:lagged} and \ref{fig2app:constant}).

\subsection{Event-time effect results}

\begin{figure}[h!]
\centering
\includegraphics[scale=0.28]{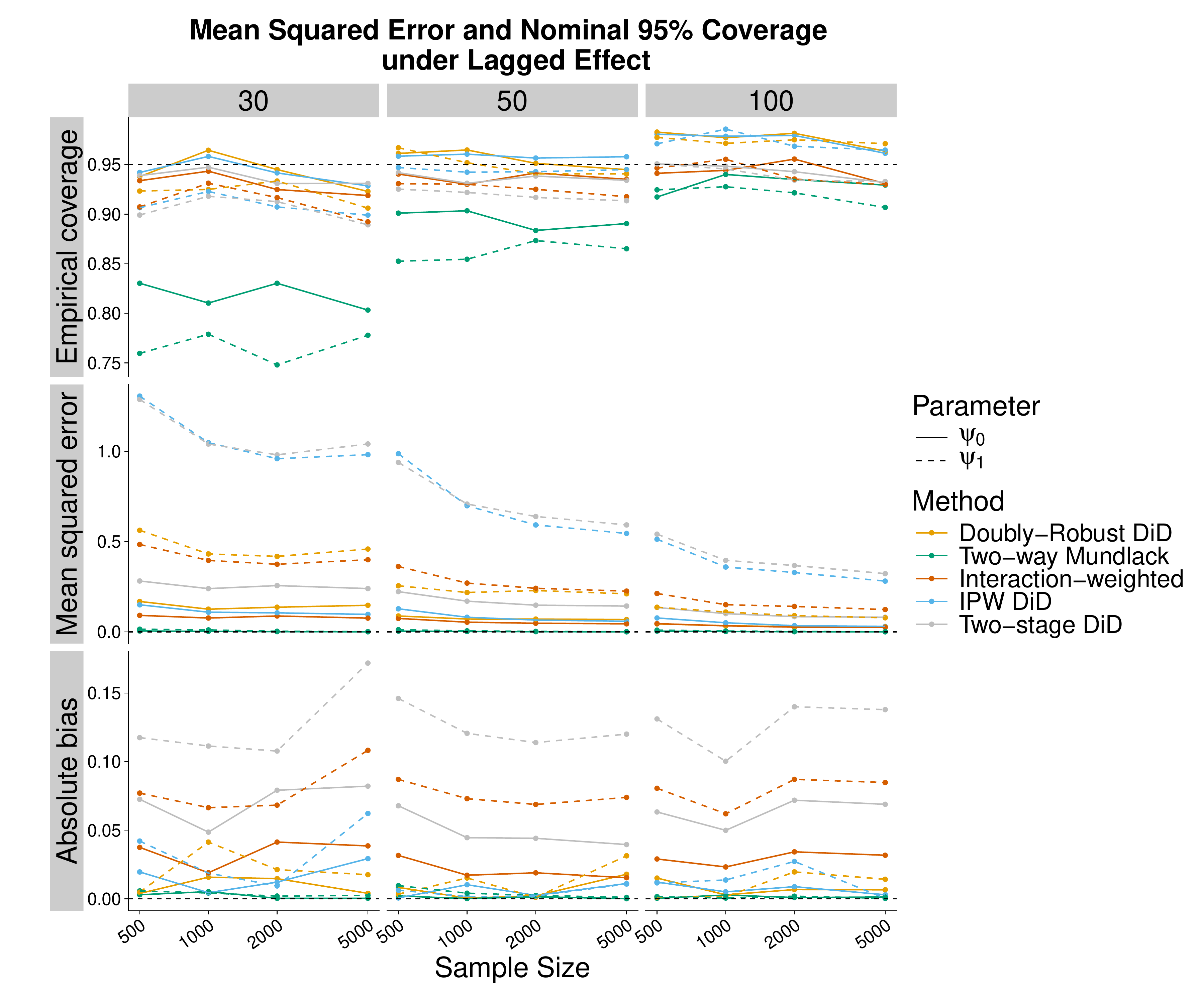}
\caption{Performance results for each of the two event-time estimates ($\psi_{0}$ and $\psi_{1}$) by sample size and number of clusters in Scenario 2. Top panel: empirical 95\% coverage. Middle panel: mean squared error (MSE). Bottom panel: absolute bias. Two-stage DiD and the Interaction-weighted estimators had the highest absolute bias, and did not improve as the numbers of clusters increased. The weighted DiD method for multiple time periods and Two-stage DiD had the highest MSE, while Two-way Mundlack regression had the lowest. All methods had relatively stable MSE within cluster sizes, and slightly improved as the number of clusters increased. All methods except both versions of DiD for Multiple Time Periods showed under-coverage, with Two-way Mundlack regression having the lowest coverage; however, coverage improved as the number of clusters increased.}
\label{fig3:eventtime}
\end{figure}

Figure \ref{fig3:eventtime} shows the results for the event-time parameters in Scenario 2. In general, the methods performed better when estimating $\psi_0$ compared to $\psi_1$. As with group-time effects, this difference is primarily driven by the number of control units available to estimate each parameter. 

Two-way Mundlack regression had the lowest MSE among all methods; however, its empirical coverage was substantially below 95\%, though it improved as the number of clusters increased. The Interaction-weighted estimator outperformed both the DiD for multiple time periods and the Two-stage DiD estimator in terms of MSE, while achieving comparable coverage. Nevertheless, its bias was worse compared to DiD for multiple time periods.  The methods with the highest MSE were the IPW DiD and Two-stage DiD. However, the Two-stage DiD had under-coverage --- probably due to its high bias ---  while both of the DiD for multiple time periods approaches displayed over-coverage, which is also expected due to its simultaneous confidence bands.

\subsection{Marginal effect}

\begin{figure}[h!]
\centering
\includegraphics[scale=0.28]{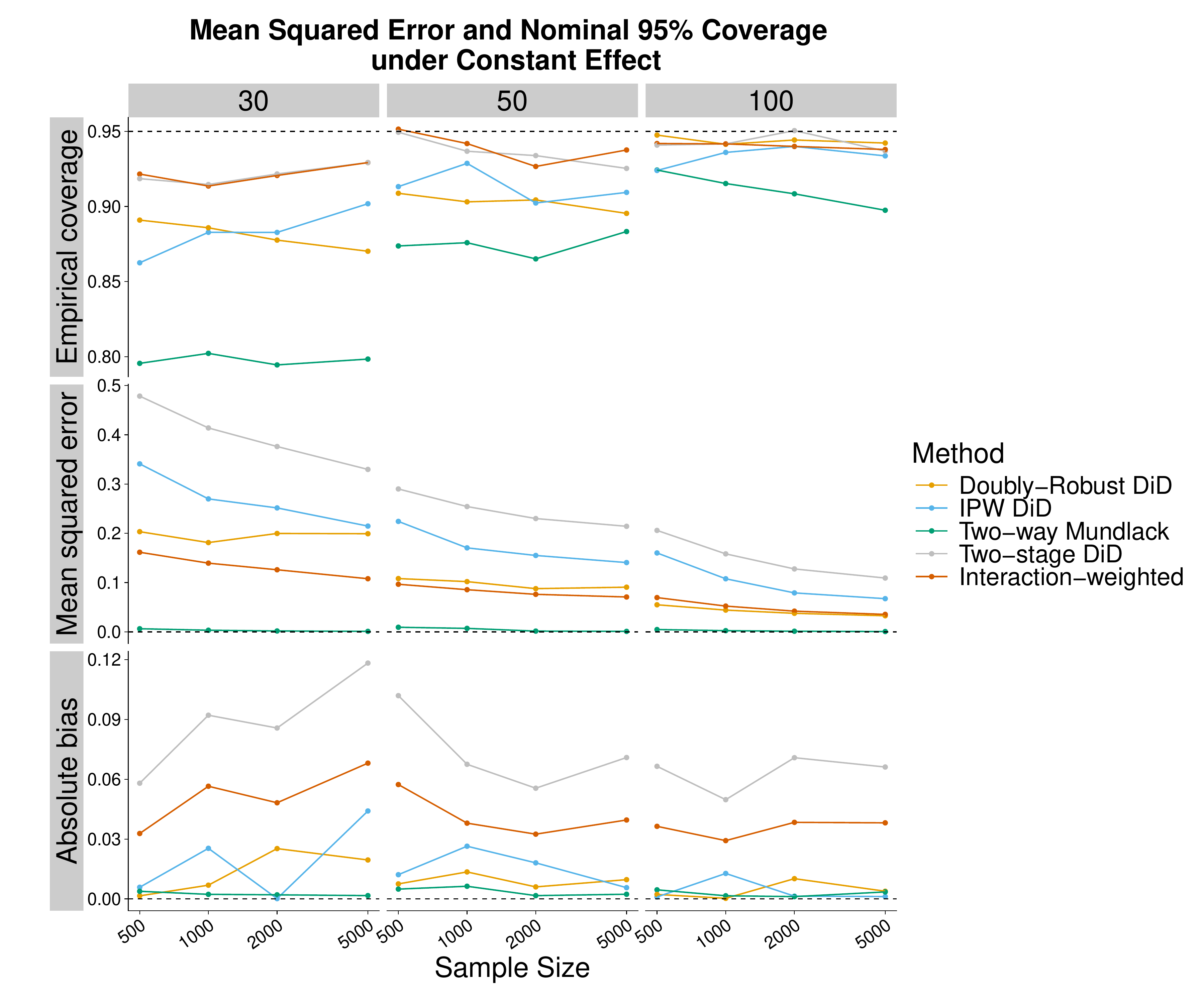}
\caption{Performance results by sample size and number of clusters for Scenario 1 when estimating $\psi_{aggr}$. Top panel: empirical 95\% coverage. Middle panel: mean squared error (MSE). Bottom panel: absolute bias. Two-stage DiD had the highest MSE, while the Two-way Mundlack regression had the lowest. Except for the doubly-robust DiD, all methods had modest to large improvements in MSE within a given number of clusters. All methods showed overall improvement in MSE as the number of clusters increased.  However, the Two-stage DiD and the Interaction-weighted estimators had substantial bias which did not attenuate with larger number of clusters. Coverage for all methods was below the nominal 95\%, but also improved as the number of clusters increased.}
\label{fig4:marginal}
\end{figure}

Figure \ref{fig4:marginal} shows the MSE and empirical coverage in the constant-effect scenario when estimating $\psi_{aggr}$. Consistent with the findings of the previous settings, all methods showed under-coverage when the number of clusters was small, with modest improvements as the number of clusters increased. The Two-stage DiD estimator achieved the closest nominal coverage overall; however, it presented the largest absolute bias and MSE. The Interaction-weighted estimator had a low MSE compared to the rest of the methods but also presented high bias. The DiD estimator for multiple time periods performed better in terms of coverage when the number of clusters was large. Its doubly-robust version had lower bias and MSE than the IPW version. Two-way Mundlack regression achieved the lowest MSE and bias, but exhibited poor coverage.

\section{HMSA Payment program evaluation}
\label{sec:HMSA}

To illustrate the performance of the methods in a real-world setting, we applied each DiD method to the HMSA dataset. Briefly, HMSA implemented a new payment system, referred to as 3PC, which replaced the traditional primary care payment model. The 3PC system introduced two primary changes. First, fee-for-service payments were replaced with risk-adjusted payments based on patient resource utilization and plan type for attributed members. Second, the program introduced a shared savings incentive: physician organizations could earn additional payments if their average risk-adjusted total member spending fell below a predefined benchmark. Additional details on the program are provided in \cite{navathe2019association}.

Our initial sample consisted of 3,824,997 observations from 630 providers and 750,939 members enrolled in HMSA sometime during the 5-year observation period (2014-2018). Each observation was at the member-year level and included a summary of healthcare utilization, demographic characteristics, and provider information for that year. To be included in the analysis, members were required to remain continuously enrolled with the same provider throughout the 5-year observation window, resulting in a balanced panel. We further restricted the sample by excluding advanced practice registered nurses, pediatrics and naturopath providers, as well as those providers whose geographic location of medical school and residency was missing.  After applying all eligibility criteria, the final analytic sample included 304 providers and a total of 144,523 members, resulting in 722,615 observations across 5 years. 

Baseline provider-level covariates, measured in 2014, included: urban versus rural location, island where they practice (Oahu, Hawaii, Maui, other), specialty type (Internal Medicine, Family Physician, other), location of medical school training (Hawaii, elsewhere in the United States, or international) and location residency (Hawaii or elsewhere in the United States). Baseline individual-level characteristics, also measured in 2014, included: sex, age (in years), and episode risk group (ERG) health assessment score, where a higher ERG score indicates higher comorbitity. 

The exposure was the adoption of the 3PC system. The 3PC system was adopted in a staggered manner by providers, starting in 2016. The untreated periods were 2014 and 2015. Once adopted, providers continued to use the system for the rest of the observation period. The outcome of interest was the total number of primary care visits per member per year. The program was expected to reduce number of primary care visits, in part because it incentivized the use of telehealth services. We estimated the effect of the intervention on the annual number of primary care provider (PCP) visits per member. Specifically, the target parameters included group time effects $\psi_{2016,2016}$, $\psi_{2017,2017}$, and $\psi_{2016,2017}$; event-time parameters $\psi_{0}$ and $\psi_{1}$; and the overall program effect $\psi_{aggr}$.

Descriptive statistics for both individual- and provider-level characteristics are presented in Table \ref{tab:descrip}. The number of members was distributed approximately equally across years where their associated provider joined the program. We observed some heterogeneity in provider and member characteristics with respect to group. Provider-level information had high variation with respect to medical training and location. Those who attended Medical School in Hawaii as well as their residency had a higher percentage of joining the 3PC program early. In addition, urban locations tended to join the intervention earlier compared to their rural counterparts. With respect to member covariates, we observed an average baseline ERG score that was lower in earlier years, while age,  sex, and their insurance type were mostly even across groups. 

\begin{table}[h!]
\begin{tabular}{lllllll}
& & & \multicolumn{3}{c}{Year of Program Entry}& \\ \hline
& Variable& Category & \multicolumn{1}{c}{\begin{tabular}[c]{@{}c@{}}2016\\ 44025 (30\%)\end{tabular}} & \multicolumn{1}{c}{\begin{tabular}[c]{@{}c@{}}2017\\ 54937 (38\%)\end{tabular}} & \multicolumn{1}{c}{\begin{tabular}[c]{@{}c@{}}2018\\ 46561(32\%)\end{tabular}} & \multicolumn{1}{c}{\begin{tabular}[c]{@{}c@{}}All members\\ 144523\end{tabular}} \\ \hline
\multirow{13}{*}{\begin{tabular}[c]{@{}l@{}}Provider\\ Information\end{tabular}} & \multirow{3}{*}{Specialty} & Other& 3035 (7\%)& 2644 (5\%)& 2002 (4\%)& 7681 (5\%) \\
& & \begin{tabular}[c]{@{}l@{}}Internal \\ Medicine\end{tabular} & 33185 (75\%) & 39874 (73\%) & 32861 (71\%)& 105920 (73\%) \\
& & \begin{tabular}[c]{@{}l@{}}Family\\ Physician\end{tabular} & 7805 (18\%)& 12419 (23\%) & 11698 (25\%)& 31922 (22\%)\\ \cline{2-7} 
& \multirow{3}{*}{\begin{tabular}[c]{@{}l@{}}Medical \\ School\end{tabular}} & International& 5492 (12\%)& 10514 (19\%) & 5960 (13\%) & 21966 (15\%)\\
& & Hawaii & 28275 (64\%) & 30183 (55\%) & 20455 (44\%)& 78913 (54\%)\\
& & Other& 10258 (23\%) & 14240 (26\%) & 20146 (43\%)& 44644 (31\%)\\ \cline{2-7} 
& \begin{tabular}[c]{@{}l@{}}Urban \\ or rural \\ location\end{tabular}& Urban& 41844 (95\%) & 38366 (70\%) & 35814 (77\%)& 116024 (80\%) \\ \cline{2-7} 
& \multirow{4}{*}{Island}& Oahu & 41226 (94\%) & 38520 (70\%) & 32935 (71\%)& 112681 (77\%) \\
& & Hawaii & 123 ($<1$\%) & 15222 (28\%) & 1945 (4\%)& 17290 (12\%)\\
& & Other& 490 (1\%) & 557 (1\%) & 8346 (18\%) & 9393 (6\%) \\
& & Maui & 2186 (5\%)& 638 (1\%) & 3335 (7\%)& 6159 (4\%) \\ \cline{2-7} 
& \multirow{2}{*}{\begin{tabular}[c]{@{}l@{}}Location of\\ residency\end{tabular}} & Other US & 20981 (48\%) & 28521 (52\%) & 30556 (66\%)& 80058 (55\%)\\
& & Hawaii & 23044 (52\%) & 26416 (48\%) & 16005 (34\%)& 65465 (45\%)\\ \hline
\multirow{6}{*}{\begin{tabular}[c]{@{}l@{}}Member\\ Information\end{tabular}}& \multirow{3}{*}{\begin{tabular}[c]{@{}l@{}}Line of \\ Business\end{tabular}} & Quest& 2329 (5\%)& 4304 (8\%)& 2106 (5\%)& 8739 (6\%) \\
& & Commerical& 37471 (85\%) & 45095 (82\%) & 40285 (87\%)& 122851 (84\%) \\
& & Akamai & 4225 (10\%)& 5538 (10\%)& 4170 (9\%)& 13933 (10\%)\\ \cline{2-7} 
& Sex& \begin{tabular}[c]{@{}l@{}}Prop \\ female\end{tabular} & 0.52& 0.53& 0.53 & 0.53 \\ \cline{2-7} 
& \begin{tabular}[c]{@{}l@{}}Age in\\ years\end{tabular} & mean (sd)& 55.05 (17.22) & 54.42 (18)& 54.98 (17.64)& 54.79 (17.65)\\ \cline{2-7} 
& \begin{tabular}[c]{@{}l@{}}ERG \\ score\end{tabular} & mean (sd)& 1.86 (2.77) & 1.93 (2.84) & 1.94 (2.88)& 1.91 (2.83)\\ \hline
\end{tabular}
\caption{Member characteristics by year of program adoption, based on the timing of their provider's adoption. Provider-level characteristics varied across adoption years. Notably, providers who attended medical school in Hawaii, completed residency in Hawaii, or offered services in urban areas were more likely to adopt the program earlier. At the member level, average ERG score is larger in later years of adoption year. The rest of the member-level covariates did not vary substantially across groups. }
\label{tab:descrip}
\end{table}

\subsection{Estimated effects of 3PC program}

\begin{figure}[h!]
\centering
\includegraphics[scale=0.35]{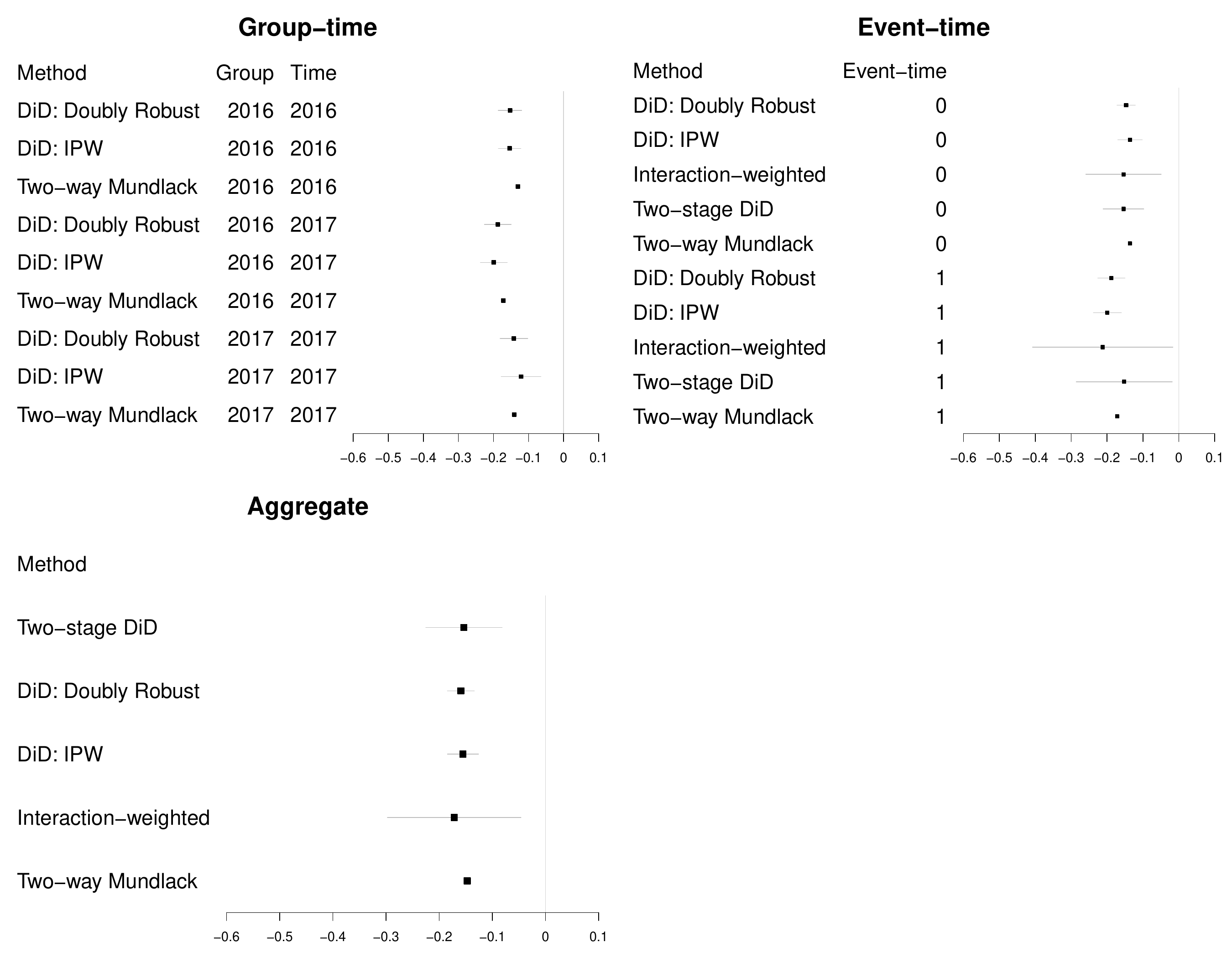}
\caption{Results from fitting the DiD estimators in the HMSA data. All methods estimate a reduction in annual primary care visits. Group-time results were similar pointwise, with DiD: IPW having largest 95\% confidence intervals, and Two-way Mundlack the smallest. Event time results indicated slightly stronger effects for $\psi_{1}$ compared to $\psi_0$. All methods had similar estimated aggregated effects.}
\label{fig:hmsaresults}
\end{figure}

We estimated the effects of joining the 3PC program using all the staggered adoption methods considered in this study, incorporating the full set of covariates and accounting for clustering at the provider level. The estimated parameters for each estimand and method are presented in Figure \ref{fig:hmsaresults}, with specific values provided in 
Tables \ref{tab:HMSAgrouptime}, \ref{tab:HMSAaggr}, and \ref{tab:HMSAeventtime}.

At the group-time level, all methods estimated a reduction in annual PCP visits per member. The estimated effects for $\psi_{2016,2016}$ ranged between 0.1 and 0.2 visit reductions per year, with 95\% confidence intervals that excluded zero. Within each method, the estimates for $\psi_{2016,2017}$ were slightly lower than those for $\psi_{2016,2016}$, indicating a greater reduction in PCP visits, suggesting a potential strengthening of the program’s effect over time for that group. Similarly, estimates for $\psi_{2017,2017}$ showed reductions in PCP visits, with effects comparable to those of $\psi_{2016,2016}$. Consistent with our simulations, the Difference-in-Differences (DiD) method for multiple time periods produced narrower confidence intervals for $\psi_{2016,2016}$ than for $\psi_{2016,2017}$.

Event-time parameters reflected reductions in PCP visits across all methods. Within each method, the reduction for $\psi_{1}$ was slightly smaller than for $\psi_{0}$, suggesting a continued improvement over time.

Aggregated estimates also showed a reduction in annual PCP visits across all methods. The IPW DiD estimator had slightly higher variance compared to its doubly robust counterpart, while the interaction-weighted estimator had the largest standard errors across its estimands. This may be due to the exclusion of observations indexed in 2018, as required in the absence of a never-treated group \cite{sun2021estimating}. The two-stage DiD method also produced wider confidence intervals than DiD for multiple time periods. Finally, the Two-way Mundlack approach produced the narrowest confidence intervals which may lead to poor coverage per our simulation results.

\section{Discussion}
\label{sec:disc}

In this study, we provided a theoretical and practical overview of four recently developed DiD methods for staggered adoption interventions. The methods share common assumptions, including limited anticipation, irreversibility of treatment, and overlap. However, they differ in their parallel trends assumptions and how they estimate nuisance parameters.

From a theoretical perspective, DiD for Multiple Time Periods relies on time- and group-specific conditional parallel trends assumptions. The Interaction-Weighted estimator also relies on similar assumptions, but requires the stricter condition of no treatment anticipation. Two-way Mundlack employs a parallel trends assumption specific to its model structure and does not allow for treatment anticipation. Two-stage DiD uses an intuitive approach to impute counterfactuals using a parallel trends assumption similarly strong to the one employed by Two-stage DiD. While the parallel trends assumptions are unverifiable with observational data, one can assess potential violations of parallel trends. Current recommendations include evaluating pre-intervention effects, such as group-time effects at negative times \cite{wang2024advances} (eg., group-time estimand $\psi_{2016,2015}$ or event-time estimand $\psi_{-1}$ in our HMSA example). Except for Two-way Mundlack, all methods allow for pre-intervention effect estimation, which can help indicate if there is potential bias if methods estimate a non-null effect. More formal hypothesis tests can be conducted, such as the method proposed by \cite{rambachan2023more}, which constructs confidence intervals to bound the extent of post-treatment violations using pre-treatment differences in trends. Currently, this approach is available for DiD with Multiple Time Periods \cite{honestdidRpkg}. 

Regarding estimation of nuisance functions, all methods except DiD for Multiple Time Periods using IPW rely on an outcome regression based on covariates. The doubly robust DiD for multiple time periods approach is appealing due to its robustness property, providing consistent estimates even if one of the nuisance models is misspecified. However, this method may struggle with sparse covariate distributions. For example, if no providers in the group 2018 had attended medical school in Hawaii, the parameter $\psi_{2017, 2017}$ could not have been estimated. Two-way Mundlack avoids this issue by using one linear model which borrows more information across groups, and while it achieved greater efficiency by fitting a single regression model, it comes at the cost of model complexity, including transformed baseline covariates that may be challenging to interpret. The Interaction-Weighted estimator may be preferable to practitioners as directly modeling the outcome as a function of covariates and event-time indicators may be more interpretable. Two-stage DiD is advantageous when counterfactual imputation is of interest and provides an intuitive approach to remove trends from the untreated group.

The choice of control groups is also crucial, affecting both interpretation and the plausibility of the identification assumptions. In the HMSA example, providers who joined the program later differed significantly from those who joined earlier. We recommend selecting methods that define controls transparently, ensuring clear interpretation and credible parallel trend assumptions. Not-yet-treated units may reduce confounding but should be used cautiously in the presence of anticipatory effects, which can bias estimates. Conversely, never-treated units are not subject to anticipation bias but may differ substantially from treated units, potentially violating the parallel trends assumption \cite{wang2024advances}. 

Multiple estimands arise in the staggered setting in order to address group and time heterogeneity. The target estimand should be prespecified and align with the scientific question of interest. For group-time effects, Two-way Mundlack produced unbiased estimates in our simulations and had lowest MSE. However, it produced under-coverage, likely due to software limitations in handling clustering in the marginalization step. DiD with multiple time periods exhibited better coverage of group-time effects, with over-coverage as the number of clusters increased. Additionally, because fewer controls are available, estimation for $\psi_{g,t}$ was more precise in the simulations when $g$ and $t$ were small regardless of the method. 

Aggregated effects require careful attention when selecting them as the target estimand due to their interpretation, and various weighting schemes that depend on the method being used. When event-time effects are of interest, one can consider utilizing the interaction-weighted estimator as it performed reasonably well overall. Two-stage DiD showed higher MSE and bias in both of the marginal effects and, importantly, its weighting scheme for estimating event-time effects is unclear. DiD for multiple time periods showed over-coverage when targeting event-time estimates due to its conservative standard errors. However, this estimator had under-coverage when estimating the overall treatment effect. The Two-way Mundlack regression, while efficient, lacks clear weighting and produced poor coverage in our simulations. In a similar fashion as group-time estimates, the MSE for $\psi_{0}$ was smaller than $\psi_{1}$, highlighting the importance of sufficient control observations for accurate estimation and power calculations. 

\begin{figure}[h!]
\centering
\includegraphics[scale=0.55]{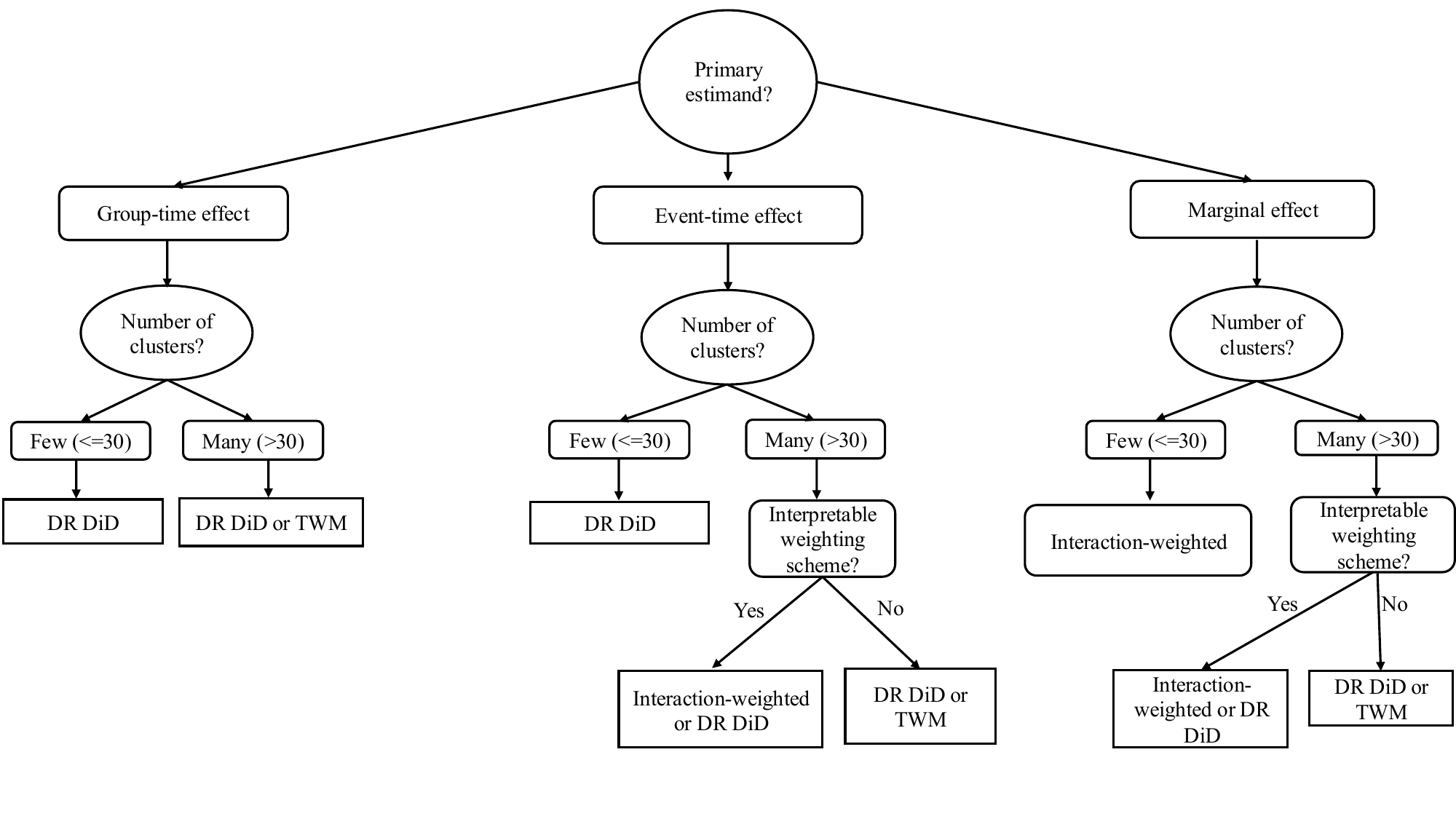}
\caption{Flowchart to help select a staggered-adoption DiD method based on estimand, number of clusters, and interpretability assuming the data analyzed satisfies the parallel trends assumption each method requires. Abbreviations: DR DiD: Difference-in-Differences for Multiple Time Periods with doubly robust estimation; TWM: Two-way Mundlack regression.}
\label{fig:flowchart}
\end{figure}

\begin{figure}[h!]
\centering
\includegraphics[scale=0.23]{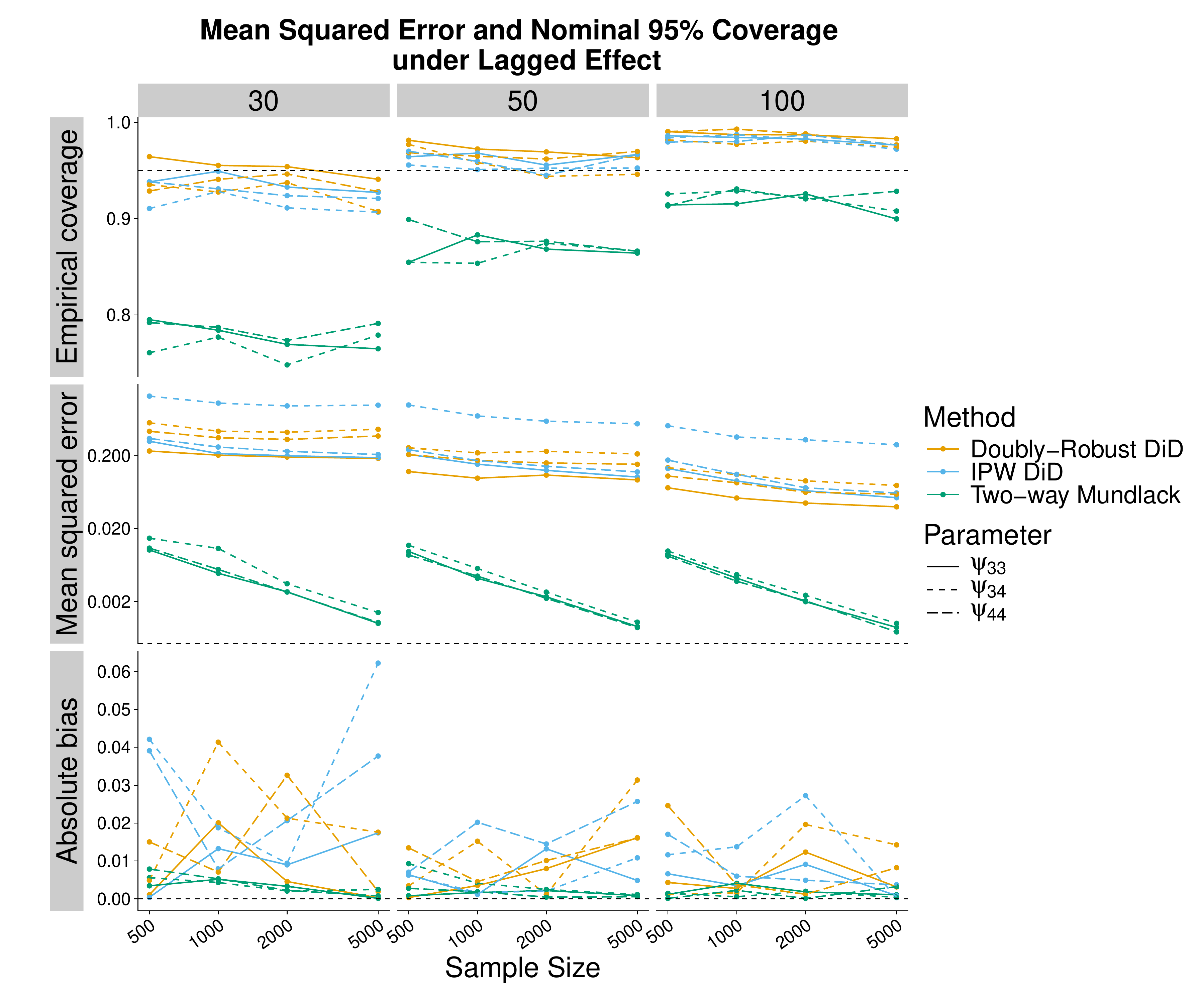}
\caption{Top panel: mean squared error for each of the 3 parameter estimates by sample size and cluster size in Scenario 2. Lower panel: empirical 95\% coverage by sample size and cluster size. }
\label{fig1app:lagged}
\end{figure}

\begin{figure}[h!]
\centering
\includegraphics[scale=0.23]{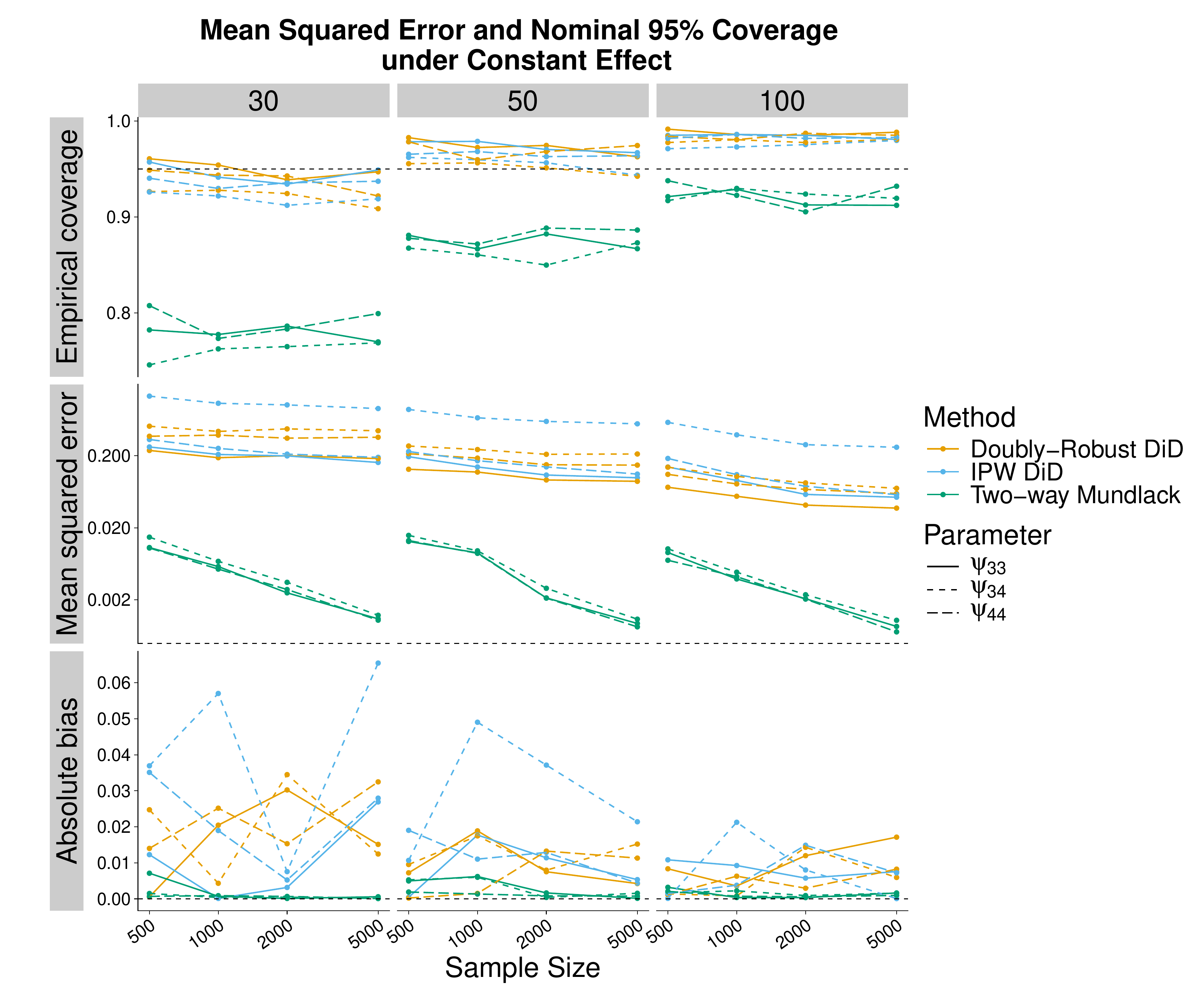}
\caption{Top panel: mean squared error for each of the 3 parameter estimates by sample size and cluster size in Scenario 1. Lower panel: empirical 95\% coverage by sample size and cluster size.}
\label{fig2app:constant}
\end{figure}

To translate our findings into practical guidance, we developed a decision flowchart (Figure~\ref{fig:flowchart}) to assist researchers in selecting an appropriate DiD method for staggered adoption settings. The flowchart organizes recommendations by estimand of interest, number of clusters, and, for marginal effects, whether an interpretable weighting scheme is desired. The first decision node identifies the estimand of interest. Because our simulations showed that performance varied with the number of clusters, the next decision layer provides guidance based on this factor. For marginal estimands, an additional node considers the importance of weighting scheme interpretability. Each recommended method is presented with its performance trade-offs. When two methods are suggested, this reflects a balance between coverage, bias, and MSE. For example, with many clusters, Two-way Mundlack achieves the lowest MSE but suffers from under-coverage, whereas DiD with Multiple Time Periods generally has over-coverage and larger MSE. Thus, for group-time effects, either may be chosen when many clusters are available. The same logic applies to marginal effects if one does not require an interpretable weighting scheme; otherwise, the Interaction-Weighted or Doubly-Robust DiD estimators are preferable. With fewer clusters, the Doubly-Robust DiD approach tends to perform better for group-time and event-time effects, while the Interaction-Weighted estimator performs better for marginal effects. In contrast, the Two-stage DiD often exhibited substantial bias and high MSE, and the IPW DiD also showed large MSE, so we do not recommend these methods in this setting.

Our simulation study has certain limitations. First, we assumed equal sample sizes across clusters, which may not reflect real-world settings. If cluster size is correlated to the intervention's effect and with the staggered design (e.g., larger clusters being treated earlier and having stronger effects), then the methods may fail to accurately estimate the treatment effect. Work has been carried out to address this issue in step-wedge trials \cite{wang2024model}. This scenario would be valuable to explore in future work in a setting without randomization. Another direction for further investigation is the impact of treatment anticipation, or switching. In our simulation study, we assumed that there was no treatment anticipation and no treatment switching. Similarly, in the HMSA analysis, we restricted the sample to individuals who did not switch providers to ensure treatment consistency (i.e., to avoid members switching from treated to untreated providers). Finally, future research could examine alternative methods not considered in this study, such as those proposed by \cite{de2020two}, which can accommodate treatment switching.

\section{Appendix}

\begin{table}[h!]
\centering
\begin{tabular}{llll}
Method               & Scenario & Parameter     & Average of estimated values \\ \hline
Doubly-Robust DiD    & 1 & $\psi_{aggr}$ & 2.00 \\
IPW DiD              & 1 & $\psi_{aggr}$ & 2.00 \\
Interaction-weighted & 1 & $\psi_{aggr}$ & 1.96 \\
Two-stage DiD        & 1 & $\psi_{aggr}$ & 1.93 \\
Two-way Mundlack     & 1 & $\psi_{aggr}$ & 2.00 \\
Doubly-Robust DiD    & 2 & $\psi_{aggr}$ & 0.63 \\
IPW DiD              & 2 & $\psi_{aggr}$ & 0.62 \\
Interaction-weighted & 2 & $\psi_{aggr}$ & 0.57 \\
Two-stage DiD        & 2 & $\psi_{aggr}$ & 0.53 \\
Two-way Mundlack     & 2 & $\psi_{aggr}$ & 0.63 \\
Doubly-Robust DiD    & 3 & $\psi_{aggr}$ & 1.68 \\
IPW DiD              & 3 & $\psi_{aggr}$ & 1.67 \\
Interaction-weighted & 3 & $\psi_{aggr}$ & 1.63 \\
Two-stage DiD        & 3 & $\psi_{aggr}$ & 1.59 \\
Two-way Mundlack     & 3 & $\psi_{aggr}$ & 1.68 \\
Doubly-Robust DiD    & 1 & $\psi_{0}$  & 2.00 \\
IPW DiD              & 1 & $\psi_{0}$  & 2.00 \\
Interaction-weighted & 1 & $\psi_{0}$  & 2.00 \\
Two-stage DiD        & 1 & $\psi_{0}$  & 2.01 \\
Two-way Mundlack     & 1 & $\psi_{0}$  & 2.00 \\
Doubly-Robust DiD    & 1 & $\psi_{1}$  & 2.01 \\
IPW DiD              & 1 & $\psi_{1}$  & 2.02 \\
Interaction-weighted & 1 & $\psi_{1}$  & 2.02 \\
Two-stage DiD        & 1 & $\psi_{1}$  & 2.03 \\
Two-way Mundlack     & 1 & $\psi_{1}$  & 2.00 \\
Doubly-Robust DiD    & 2 & $\psi_{0}$  & 0.01 \\
IPW DiD              & 2 & $\psi_{0}$  & 0.00 \\
Interaction-weighted & 2 & $\psi_{0}$  & -0.0 \\
Two-stage DiD        & 2 & $\psi_{0}$  & -0.0 \\
Two-way Mundlack     & 2 & $\psi_{0}$  & 0.00 \\
Doubly-Robust DiD    & 2 & $\psi_{1}$  & 2.01 \\
IPW DiD              & 2 & $\psi_{1}$  & 2.00 \\
Interaction-weighted & 2 & $\psi_{1}$  & 1.92 \\
Two-stage DiD        & 2 & $\psi_{1}$  & 1.96 \\
Two-way Mundlack     & 2 & $\psi_{1}$  & 2.00 \\
Doubly-Robust DiD    & 3 & $\psi_{0}$  & 1.53 \\
IPW DiD              & 3 & $\psi_{0}$  & 1.53 \\
Interaction-weighted & 3 & $\psi_{0}$  & 1.50 \\
Two-stage DiD        & 3 & $\psi_{0}$  & 1.46 \\
Two-way Mundlack     & 3 & $\psi_{0}$  & 1.54 \\
Doubly-Robust DiD    & 3 & $\psi_{1}$  & 2.00 \\
IPW DiD              & 3 & $\psi_{1}$  & 1.99 \\
Interaction-weighted & 3 & $\psi_{1}$  & 1.92 \\
Two-stage DiD        & 3 & $\psi_{1}$  & 1.87 \\
Two-way Mundlack     & 3 & $\psi_{1}$  & 2.00 \\ \hline
\end{tabular}
\label{tab:avgestimates}
\caption{Average of event time and marginal estimates across methods and scenarios when the number of observations is 5,000 and the number of clusters is 100.}
\end{table}

\begin{table}[h!]
\centering
\begin{tabular}{lllccc}
Method & Group    & Time   & Estimated effect & Estimated Sd & 95\%CI \\ \hline
IPW DiD & 2016 & 2015 & 0.04  & 0.02 & (0.00 ,0.07) \\
IPW DiD & 2016 & 2016 & -0.15 & 0.02 & (-0.19,-0.12) \\
IPW DiD & 2016 & 2017 & -0.20 & 0.02 & (-0.24,-0.16) \\
IPW DiD & 2017 & 2015 & -0.01 & 0.02 & (-0.06,0.04) \\
IPW DiD & 2017 & 2016 & -0.06 & 0.03 & (-0.11,0.00) \\
IPW DiD & 2017 & 2017 & -0.12 & 0.03 & (-0.18,-0.06) \\
Doubly-Robust DiD & 2016 & 2015 & 0.04  & 0.02 & (0.00 ,0.07) \\
Doubly-Robust DiD & 2016 & 2016 & -0.15 & 0.02 & (-0.19,-0.12) \\
Doubly-Robust DiD & 2016 & 2017 & -0.19 & 0.02 & (-0.23,-0.15) \\
Doubly-Robust DiD & 2017 & 2015 & 0.02  & 0.02 & (-0.01,0.06) \\
Doubly-Robust DiD & 2017 & 2016 & -0.04 & 0.02 & (-0.07,0.00) \\
Doubly-Robust DiD & 2017 & 2017 & -0.14 & 0.02 & (-0.18,-0.10) \\
Two-way Mundlack  & 2016 & 2016 & -0.13 & 0.00 & (-0.14,-0.12) \\
Two-way Mundlack  & 2016 & 2017 & -0.17 & 0.00 & (-0.17,-0.17) \\
Two-way Mundlack  & 2017 & 2017 & -0.14 & 0.00 & (-0.14,-0.14) \\ \hline
\end{tabular}
\caption{Estimated group-time effects in HMSA data set by method.}
\label{tab:HMSAgrouptime}
\end{table}

\begin{table}[h!]
\centering
\begin{tabular}{llccc}
Method & $\psi_\ell$ & Estimated effect & Estimated Sd & 95\% CI \\ \hline
IPW DiD     & $\psi_{-2}$        & -0.01 & 0.02 & (-0.06,0.03)\\
IPW DiD     & $\psi_{-1}$        & -0.01   & 0.02& (-0.04,0.02)\\
IPW DiD     & $\psi_{0 }$        & -0.14   & 0.02& (-0.17,-0.10)\\
IPW DiD     & $\psi_{1 }$        & -0.20   & 0.02& (-0.24,-0.16)\\
Doubly-Robust DiD    & $\psi_{-2}$        & 0.02    & 0.02& (-0.01,0.06)\\
Doubly-Robust DiD    & $\psi_{-1}$        & 0.00    & 0.01& (-0.03,0.02)\\
Doubly-Robust DiD    & $\psi_{0 }$        & -0.15   & 0.01& (-0.17,-0.12)\\
Doubly-Robust DiD    & $\psi_{1 }$        & -0.19   & 0.02& (-0.23,-0.15)\\
Two-way Mundlack     & $\psi_{0 }$        & -0.14   & 0.00& (-0.14,-0.13)\\
Two-way Mundlack     & $\psi_{1 }$        & -0.17   & 0.00& (-0.17,-0.17)\\
Two-stage DiD        & $\psi_{-4}$        & 0.02    & 0.02& (-0.01,0.05)\\
Two-stage DiD        & $\psi_{-3}$        & -0.01   & 0.01& (-0.03,0.01)\\
Two-stage DiD        & $\psi_{-2}$        & 0.00    & 0.01& (-0.02,0.01)\\
Two-stage DiD        & $\psi_{0 }$        & -0.15   & 0.03& (-0.21,-0.10)\\
Two-stage DiD        & $\psi_{1 }$        & -0.15   & 0.07& (-0.29,-0.02)\\
Interaction-weighted & $\psi_{-3}$        & 0.12    & 0.09& (-0.06,0.30)\\
Interaction-weighted & $\psi_{-2}$        & 0.01    & 0.02& (-0.03,0.04)\\
Interaction-weighted & $\psi_{0 }$        & -0.15   & 0.05& (-0.26,-0.05)\\
Interaction-weighted & $\psi_{1 }$        & -0.21   & 0.10& (-0.41,-0.02) \\ \hline
\end{tabular}
\caption{Estimated event-time effects in HMSA data set by method.}
\label{tab:HMSAeventtime}
\end{table}

\begin{table}[h!]
\centering
\begin{tabular}{llccc}
Method & Estimated effect & Estimated Sd & 95\%CI  \\ \hline
Two-stage DiD        & -0.15            & 0.04         & (-0.23,-0.08) \\
IPW DiD              & -0.16            & 0.01         & (-0.18,-0.13) \\
Doubly-Robust DiD    & -0.16            & 0.01         & (-0.18,-0.13) \\
Two-way Mundlack     & -0.15            & 0.001           & (-0.15,-0.14) \\
Interaction-weighted & -0.17            & 0.06         & (-0.3 ,-0.05) \\ \hline
\end{tabular}
\caption{Estimated aggregated effects in HMSA data set by method.}
\label{tab:HMSAaggr}
\end{table}

\begin{table}[]
\centering
\begin{tabular}{llll}
\hline
Period     & 3    & 4    & 5    \\\hline
Percentage & 22\% & 35\% & 43\% \\
\hline
\end{tabular}
\caption{Marginal probabilities ($P(G=g)$) for each period $g$}
\label{tab:margprobs}
\end{table}

\begin{table}[]
\centering
\begin{tabular}{l|l|l|l}
              & k = 3 & k=4   & k=5   \\ \hline
$\theta_{1k}$ & -0.87 & -0.70 & -0.46 \\
$\theta_{2k}$ & 1.20  & 0.10  & 0.65  \\
$\theta_{3k}$ & 1.54  & 1.26  & 0.83  \\
$\theta_{4k}$ & 1.90  & 1.54  & 1.02  \\
$\theta_{5k}$ & -2.57 & -2.10 & -1.39 \\ \hline
\end{tabular}
\caption{Coefficients for generating program entry probabilities at each time point $k$. For each time $k$, $\theta_{1k}$ represents the intercept's coefficient, $\theta_{2k}$ to $\theta_{4k}$ are the coefficients that correspond to cluster level variable $X_4$, and $\theta_{5k}$ is the coefficient for $X_5$.}
\label{tab:coefssim}
\end{table}

\newpage 
\bibliographystyle{plain}
\bibliography{bibliography.bib}

\end{document}